\documentclass[journal,11pt,draftclsnofoot,onecolumn]{IEEEtran}

\IEEEoverridecommandlockouts                              

\usepackage{graphicx}
\usepackage[square,comma,numbers,sort&compress]{natbib}
\usepackage{amsthm,amssymb,amscd,amssymb,amsmath,mathrsfs}
\usepackage{epstopdf}
\usepackage{balance}
\usepackage[noend]{algorithm} 
\usepackage{algorithmicx} 
\usepackage[noend]{algpseudocode}
\usepackage{multirow} 
\usepackage{amsmath}
\usepackage{xcolor}
\usepackage{lineno}
\usepackage{enumerate}

\makeatletter

\newcommand{\Rmnum}[1]{\expandafter\@slowromancap\romannumeral #1@}
\makeatother

\theoremstyle{plain}
\newtheorem{theorem}{Theorem}

\theoremstyle{definition}

\theoremstyle{remark}

\begin{document}
%
\title{Supervisor Localization for Large-Scale Discrete-Event Systems under Partial Observation*
 }

\author{Renyuan Zhang$^{1}$, Kai Cai$^{2}$
\thanks{*This work was supported in part by the National Nature Science Foundation of China, Grant no. 61403308;
the Natural Science Foundation of Shaanxi Province, China, Grant no. 2017JM5061;
JSPS KAKENHI Grant no. JP16K18122.
}
\thanks{$^{1}$R. Zhang is with School of Automation, Northwestern Polytechnical University, China. Email:
        {\tt\small ryzhang@nwpu.edu.cn}.}%
\thanks{$^{2}$K. Cai is with Department of Electrical and Information Engineering, Osaka City University, Japan. Email:
        {\tt\small kai.cai@eng.osaka-cu.ac.jp}.}%
}

\maketitle

\thispagestyle{empty} \pagestyle{plain}


\begin{abstract}
Recently we developed \emph{partial-observation supervisor localization}, a top-down
approach to distributed control of discrete-event systems (DES) under partial
observation. Its essence is the decomposition of the
\emph{partial-observation monolithic supervisor} into \emph{partial-observation local controllers}
for individual controllable events.
In this paper we extend the partial-observation supervisor
localization to large-scale DES, for which the monolithic supervisor
may be incomputable. Specifically, we first employ an efficient
heterarchical supervisor synthesis procedure to compute
a heterarchical array of \emph{partial-observation
decentralized supervisors} and \emph{partial-observation coordinators}. Then we localize each
of these supervisors/coordinators into \emph{partial-observation local controllers}. This procedure suggests
a systematic approach to the distributed control of large-scale DES under partial observation.  The results
are illustrated by a system of automatic guided vehicles (AGV) serving a manufacturing
workcell.
\end{abstract}


\section{Introduction}


Recently we developed in \cite{ZhangCW17}
a top-down approach, called {\it partial-observation supervisor localization}, to the
distributed control of multi-agent discrete-event systems (DES) under partial
observation. Specifically, we first synthesize a
{\it partial-observation monolithic supervisor} using the concept of {\it relative
observability} in \cite{CaiZW15,CaiZW16}, and then
decompose the supervisor into local controllers for individual controllable events,
by a {\it partial-observation localization procedure} adapted from \cite{CaiWon10a}.
The derived local controllers have state transitions triggered only by observable
events, and they collectively achieve the same controlled behavior
as the partial-observation monolithic supervisor does.
This approach, however, cannot deal with large-scale system,
because the monolithic supervisor synthesis at the first
step is NP-hard \cite{GohWon00}; indeed the state size
of the supervisor grows exponentially in the number of
individual plant components and specifications.

In this paper, we propose a systematic attack to distributed
control of large-scale DES under partial-observation. Just
as in \cite{CaiWon10a,CaiWon10b} for full-observation case,
we combine the partial-observation supervisor localization \cite{ZhangCW17}
with an efficient heterarchical supervisor synthesis procedure \cite{FenWon08}.
Specifically, we first compute a heterarchical array of {\it partial-observation
decentralized supervisors} and {\it partial-observation coordinators} to achieve globally
feasible and nonblocking controlled behavior. In computing these
decentralized supervisors/coordinators, we (again) employ relative
observability since it is closed under set unions and the supremal sublanguage
exists. We then localize each of these partial-observation
supervisors/coordinators into partial-observation local controllers
by the partial-observation localization procedure in \cite{ZhangCW17}.
As in \cite{ZhangCW17}, the partial-observation local controllers
have only observable events causing state changes.

The contributions of this work are twofold. First, from a theoretical view,
the combination of partial-observation supervisor localization procedure with
the heterarchical supervisor synthesis procedure supplies a systematic
approach to the distributed control of large-scale discrete-event systems under
partial observation. The heterarchical supervisor synthesis procedure makes the
localization procedure efficient and thus applicable to large systems.
By employing relative observability, the derived
controlled behavior will be generally more permissive than that
derived by normality; the latter is widely used in the literature.
Second, from a practical view, this work suggests an effectively computable
way to design a distributed control architecture under partial
observation for a multi-agent plant with large size and a decomposable
specification; all the procedures are implemented by computer algorithms
(in the software package TCT \cite{Wonham16b}). The detailed steps are illustrated
by an AGV example, in which all the computations are executed by TCT procedures.

The paper is organized as follows. Section \ref{sec:prelim} reviews the supervisory
control problem of DES under partial observation and formulates the
partial-observation supervisor localization problem. Section \ref{sec:locproc}
presents the partial-observation localization procedure
for large-scale system. Section \ref{sec:casestudy} describes the AGV,
and presents the solution to the distributed control of AGV under partial observation.
Finally Section \ref{sec:concl} states our conclusions.


\section{Preliminaries and Problem Formulation} \label{sec:prelim}

\subsection{Preliminaries on Partial Observation}


The plant to be controlled is modelled by a generator
\begin{align} \label{eq:plantG}
{\bf G} = (Q,\Sigma, \delta, q_0, Q_m)
\end{align}
where $Q$ is the finite state set; $q_0 \in Q$ is the initial state;
$Q_m \subseteq Q$ is the subset of marker states; $\Sigma$ is the finite event set;
$\delta: Q\times \Sigma\rightarrow Q$ is the (partial) state transition function.
In the usual way, $\delta$ is extended to $\delta:Q\times\Sigma^*\rightarrow Q$,
and we write $\delta(q,s)!$ to mean that $\delta(q,s)$ is defined.
Let $\Sigma^*$ be the set of all finite strings, including the empty string $\epsilon$.
The {\it closed behavior} of $\bf G$ is the language
\[L({\bf G}) = \{s\in \Sigma^*|\delta(q_0,s)!\}\]
and the {\it marked behavior} is
\[L_m({\bf G}) = \{s\in L({\bf G})|\delta(q_0,s)\in Q_m\}\subseteq L({\bf G}).\]

For supervisory control, the event set $\Sigma$ is partitioned into
$\Sigma_c$, the subset of controllable events that can be disabled
by an external supervisor, and $\Sigma_{uc}$, the subset of
uncontrollable events that cannot be prevented from occurring (i.e.
$\Sigma = \Sigma_c\dot\cup\Sigma_{uc}$). For partial observation,
$\Sigma$ is partitioned into $\Sigma_o$, the subset of observable
events, and $\Sigma_{uo}$, the subset of unobservable events (i.e.
$\Sigma = \Sigma_o \dot\cup \Sigma_{uo}$). Bring in the natural
projection $P: \Sigma^* \rightarrow \Sigma_o^*$
defined by
\begin{equation} \label{eq:natpro}
\begin{split}
P(\epsilon) &= \epsilon; \\
P(\sigma) &= \left\{
  \begin{array}{ll}
    \epsilon, & \hbox{if $\sigma \notin \Sigma_o$,} \\
    \sigma, & \hbox{if $\sigma \in \Sigma_o$;}
  \end{array}
\right.\\
P(s\sigma) &= P(s)P(\sigma),\ \ s \in \Sigma^*, \sigma \in \Sigma
\end{split}
\end{equation}
As usual, $P$ is extended to $P: Pwr(\Sigma^*) \rightarrow Pwr(\Sigma_o^*)$, where
$Pwr(\cdot)$ denotes powerset. Write $P^{-1}: Pwr(\Sigma_o^*) \rightarrow Pwr(\Sigma^*)$
for the \emph{inverse-image function} of $P$.

A {\it supervisory control} for $\bf G$ is any map $V:L({\bf G})\rightarrow \Gamma$,
where $\Gamma := \{\gamma \subseteq \Sigma|\gamma \supseteq \Sigma_{uc}\}$. Then the
closed-loop system is $V/{\bf G}$, with closed behavior $L(V/{\bf G})$ and marked
behavior $L_m(V/{\bf G})$ \cite{Wonham16a}.
Under partial observation $P:\Sigma^*\rightarrow \Sigma_o^*$,
we say that $V$ is {\it feasible} if
\begin{equation*}
(\forall s, s' \in L({\bf G}))~ P(s) = P(s')\Rightarrow V(s) = V(s')
\end{equation*}
and $V$ is {\it nonblocking} if $\overline{L_m(V/{\bf G})} = L(V/{\bf G})$.

It is well-known \cite{LinWon88} that under partial observation, a feasible
and nonblocking supervisory control $V$ exists which synthesizes a (nonempty)
sublanguage $K\subseteq L_m({\bf G})$ if and only if $K$ is both
controllable and observable \cite{Wonham16a}. When $K$ is not observable,
however, there generally does not exist the supremal observable (and
controllable) sublanguage of $K$. Recently in \cite{CaiZW15},
a new concept of {\it relative observability} is proposed, which is stronger
than observability but permits the existence of the supremal
relatively observable sublanguage.

Formally, a sublanguage $K \subseteq L_m({\bf G})$ is {\it
controllable} \cite{Wonham16a} if
\begin{equation*}
\overline{K}\Sigma_{uc}\cap L({\bf G}) \subseteq \overline{K}.
\end{equation*}
Let $C\subseteq L_m({\bf G})$. A sublanguage $K \subseteq C$ is {\it relatively observable} with
respect to $C$ (or $C$-observable) if for every
pair of strings $s,s'\in \Sigma^*$ that are lookalike under $P$,
i.e. $P(s)=P(s')$, the following two conditions hold
\cite{CaiZW15}:
\begin{align}
\mbox{(i)}~ &(\forall \sigma \in \Sigma) s\sigma \in \overline{K}, s'
\in \overline{C},s'\sigma\in L({\bf G})
\Rightarrow s'\sigma \in \overline{K} \label{eq:sub1:reloberv} \\
\mbox{(ii)}~ &s \in K, s' \in \overline{C} \cap L_m({\bf G})
\Rightarrow s' \in K \label{eq:sub2:reloberv}
\end{align}
For $F\subseteq L_m({\bf G})$ write $\mathcal{CO}(F)$ for the family of
controllable and $C$-observable sublanguages of $F$. Then
$\mathcal{CO}(F)$ is nonempty (the empty language $\emptyset$
belongs) and is closed under set union; $\mathcal{CO}(F)$ has a
unique supremal element $\sup \mathcal{CO}(F)$ given by
\begin{equation*}
\sup \mathcal{CO}(F) = \bigcup\{K|K\in \mathcal{CO}(F)\}
\end{equation*}
which may be effectively computed \cite{CaiZW15}.


\subsection{Formulation of Partial-Observation Supervisor Localization Problem for Large-Scale DES}


Let the plant $\bf G$ be comprised of $N$ ($>1$) component agents
\[{\bf G}_k = (Q_k, \Sigma_k, \delta_k, q_{0,k}, Q_{m,k}), k = 1,..., N.\]
Then ${\bf G}$ is the {\it synchronous product} \cite{Wonham16a} of ${\bf G}_k$ ($k$ in the integer range $\{1,...,N\}$), denoted
as $[1,N]$,
i.e.
\begin{align} \label{eq:plant}
{\bf G} := \mathop{||}\limits_{k \in [1, N]} {\bf G}_k
\end{align}
where $||$ denotes synchronous product of generators \cite{Wonham16a}.
Here $\Sigma_k$ need not be pair-wise disjoint, and thus $\Sigma = \cup\{\Sigma_k | k \in [1,N]\}$.

The plant components are implicitly coupled through a control
specification language $E$ that imposes behavioral constraints
on $\bf G$. As in the literature (e.g. \cite{LinWon88_1,WilWon91}),
assume that $E$ is {\it decomposable} into specifications
$E_p \subseteq \Sigma_{e,p}^*$ ($p \in \mathcal{P}$, $\mathcal{P}$
an index set), where the $\Sigma_{e,p} \subseteq \Sigma$ need not be
pairwise disjoint; namely
\begin{align} \label{eq:spec}
E = \mathop{||} \limits_{p \in \mathcal{P}} ~ E_p
\end{align}
where $||$ denotes synchronous product of languages \cite{Wonham16a}.
Thus $E$ is defined over $\Sigma_e := \cup\{\Sigma_{e,p}|p \in \mathcal{P}\}$.

Considering partial-observation, let $\Sigma_o$ be the observable event set.
For the plant $\textbf{G}$ and the specification $E$
described above, let $\alpha \in \Sigma_c$ be an arbitrary controllable event,
which may or may not be observable. We say that a generator
\begin{equation*}
{\bf LOC}_\alpha =
(Y_\alpha,\Sigma_\alpha,\eta_\alpha,y_{0,\alpha},Y_{m,\alpha}),\
\Sigma_\alpha \subseteq \Sigma_o \cup \{\alpha\}
\end{equation*}
is a {\it partial-observation local controller} for $\alpha$ if (i)
${\bf LOC}_\alpha$ enables/disables only the event $\alpha$, and (ii) if $\sigma$ is
unobservable, i.e. $\sigma \in \Sigma_{uo}$, then $\sigma$-transitions are selfloops in ${\bf
LOC}_\alpha$, i.e.
\begin{align*}
(\forall y \in Y_\alpha)\ \eta_\alpha(y,\sigma)! \Rightarrow
\eta_\alpha(y,\sigma) = y.
\end{align*}



Condition (i) restricts the control scope of ${\bf LOC}_\alpha$ to be
only the event $\alpha$, and condition (ii) defines the observation scope of
${\bf LOC}_\alpha$ as $\Sigma_o$.
The latter is a distinguishing feature of a partial-observation
local controller as compared to its full-observation counterpart in
\cite{CaiWon10a}; the result is that only observable events may
cause a state change in ${\bf LOC}_\alpha$, i.e.
\begin{align*} 
(\forall y, y' \in Y_\alpha, \forall \sigma \in \Sigma_\alpha)\ y' =
\eta_\alpha(y,\sigma)!,\ y' \neq y 
\Rightarrow\sigma \in \Sigma_o.
\end{align*}

Note that the event set $\Sigma_\alpha$ of ${\bf LOC}_\alpha$ in
general satisfies
\begin{equation*}
\{\alpha\} \subseteq \Sigma_\alpha \subseteq \Sigma_o \cup \{\alpha\};
\end{equation*}
in typical cases, both subset containments are strict. The
events in $\Sigma_\alpha \setminus \{ \alpha \}$ may be viewed as
communication events that are critical to achieve synchronization with other
partial-observation local controllers (for other controllable
events). The event set $\Sigma_\alpha$ is not fixed \emph{a priori}, but will
be determined as part of the localization result presented in the
next section.

We now formulate the {\it Partial-Observation Supervisor Localization
Problem}:

Construct a set of partial-observation local controllers $\{{\bf
LOC}_\alpha\ |\ \alpha \in \Sigma_c\}$ such that the collective
controlled behavior of these local controllers
is safe, i.e.
\begin{align*} 
   L_m({\bf G})\cap \Big(\mathop \bigcap\limits_{\alpha \in \Sigma_{c}}
   P_\alpha^{-1}L_m({\bf LOC}_{\alpha}) \Big) \subseteq L_m({\bf G}) \cap P_e^{-1}E
\end{align*}
and nonblocking, i.e.
\begin{align*}
   L({\bf G}) ~\cap~ \Big(\mathop \bigcap\limits_{\alpha \in \Sigma_{c}}
   P_\alpha^{-1}L({\bf LOC}_{\alpha}) \Big) 
   = \overline{L_m({\bf G}) ~\cap~ \Big(\mathop \bigcap\limits_{\alpha \in \Sigma_{c}}
   P_\alpha^{-1}L_m({\bf LOC}_{\alpha}) \Big)} 
\end{align*}
where $P_e: \Sigma ^* \rightarrow \Sigma_e^*$ and $P_\alpha : \Sigma^* \rightarrow \Sigma_\alpha^*$
are the corresponding natural projections.

Having obtained a set of partial-observation local controllers, one
for each controllable event, we can allocate each controller to the
agent(s) owning the corresponding controllable event. Thereby we
build for a multi-agent DES a nonblocking distributed control
architecture under partial observation.


\section{Partial-Observation Localization Procedure for Large-Scale DES} \label{sec:locproc}

The partial-observation supervisor localization procedure proposed
in \cite{ZhangCW17} presents a solution to the problem Partial-Observation
Supervisor Localization for small-scale DES, in which the monolithic
supervisor is assumed to be feasibly computable. The assumption may
no longer hold, however, when the system is large-scale and the problem
of state explosion arises.  In the literature, there have been several
architectural approaches proposed to deal with the computational issue
based on {\it model abstraction} \cite{HilTil06,FenWon08,SchBre11,SuSR12}.

Just as in \cite{CaiWon10a}, we propose to combine the (partial-observation)
localization procedure \cite{ZhangCW17} with an efficient heterarchical
supervisor synthesis procedure \cite{FenWon08} in an
alternative top-down manner: first synthesize a heterarchical array
of partial-observation decentralized supervisors/coordinators
that collectively achieves a globally feasible and nonblocking
controlled behavior; then apply the developed localization algorithm
to decompose each of the supervisor/coordinator into partial-observation
local controllers for the relevant controllable events.


\subsection{Localization Procedure}



Recall that we have:
\begin{itemize}

\item[-] The plant to be controlled is given by ${\bf G}$ (defined over $\Sigma$),
consisting of ${\bf G}_k$ defined over disjoint
$\Sigma_k$ ($k \in [1,N]$).

\item[-] The specification $E$ is decomposable into
$E_p \subseteq \Sigma_{e,p}^*$ ($p \in \mathcal{P}$). So
$E$ is defined over $\Sigma_e := \bigcup\{\Sigma_{e,p}|p \in \mathcal{P}\}$.

\item[-] The subset of unobservable events is $\Sigma_{uo} \subseteq \Sigma$, with
the corresponding natural projection $P :\ \Sigma^* \rightarrow \Sigma_o^*$
($\Sigma_o = \Sigma\setminus \Sigma_{uo}$).
\end{itemize}

\begin{figure}[!t]
\centering
    \includegraphics[scale = 0.80]{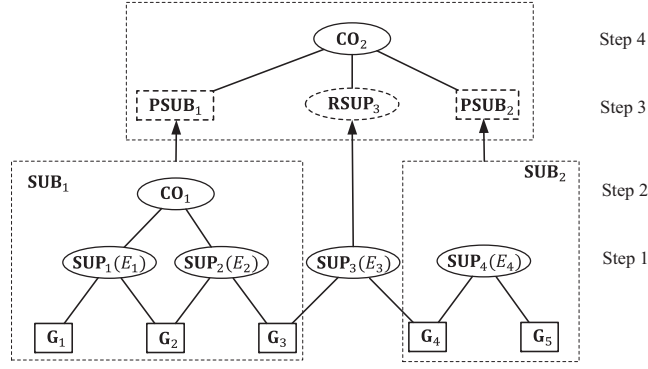}
\caption{Partial-Observation Supervisor Synthesis}
\label{fig:modsupsyn}
\end{figure}

The procedure of this partial-observation heterarchical supervisor localization
is outlined as follows; for illustration, we shall use  Fig.~\ref{fig:modsupsyn} as a running example.

\vspace{1em}
{\it Step 1) Partial-observation decentralized supervisor
synthesis}: For each control specification $E_p$ (defined on $\Sigma_p$), collect the
relevant component agents (e.g. by event-coupling), and denote
their synchronous product by ${\bf G}_p$, i.e.
\begin{align} \label{eq:decplant}
{\bf G}_p := || \{{\bf G}_k | k \in [1,N], \Sigma_k \cap \Sigma_p \neq \emptyset\}
\end{align}
Then the alphabet of ${\bf G}_p$ is
\[{\Sigma_p} := \cup \{\Sigma_k | k \in [1,N], \Sigma_k \cap \Sigma_p \neq \emptyset\}.\]
In this paper we assume that all the component agents are relevant to at least
one component specification $E_p$; thus, ${\bf G}$ is exactly the synchronous product
of all ${\bf G}_p$, i.e.
\begin{align}
L({\bf G}) &= \bigcap\limits_{p \in \mathcal{P}}P_p^{-1}L({\bf G}_p) \label{eq:sub1:compG}\\
L_m({\bf G}) &= \bigcap\limits_{p \in \mathcal{P}}P_p^{-1}L_m({\bf G}_p)\label{eq:sub2:compG}
\end{align}
Considering partial
observation $P:\Sigma^* \rightarrow \Sigma_o^*$, first compute using relative observability
a controllable and observable sublanguage
\[K_p := \sup \mathcal {CO}(E_p || L_m(\textbf{G}_p)),\]
and then construct (the construction is based on {\it uncertainty
sets} of the generator representing $K_p$ and the details are referred to \cite{ZhangCW17,Wonham16a})
a {\it partial-observation decentralized supervisor}
\begin{align}  \label{eq:decsup}
{\bf SUP}_p = (X_p,\Sigma_p,\eta_p,x_{0,p},X_{m,p})
\end{align}
such that
\begin{align*}
L_m({\bf G}_p) \cap L_m(\textbf{SUP}_p) &= K_p \\
L({\bf G}_p) \cap L(\textbf{SUP}_p) &= \overline{K_p}.
\end{align*}

This is displayed in Fig.~\ref{fig:modsupsyn}, ``Step 1" where
$E_p$ ($p = 1,2,3,4$) denotes a specification and ${\bf SUP}_p$
denotes the corresponding partial-observation decentralized supervisor.

\vspace{1em}
{\it Step 2) Subsystem decomposition and coordination}: After
Step~1, we view the system as comprised of a set of modules ${\bf M}_p (p \in \mathcal{P})$, each
consisting of a decentralized supervisor ${\bf SUP}_p$ with its associated
component agents. We decompose the system into smaller-scale
subsystems, through grouping the modules based on their
interconnection dependencies (e.g. event-coupling or control-flow
net \cite{FenWon08}).

Having obtained a set of subsystems, we verify the nonblocking
property for each of them. If a subsystem ${\bf SUB}_q$ (with event
set $\Sigma_q$) happens to be blocking, we
design a {\it partial-observation coordinator}  that removes blocking strings
\cite[Theorem~4]{FenWon08}. The design of the coordinator
must also respect partial observation $P:\Sigma^* \rightarrow \Sigma_o^*$
and the construction is similar to that of partial-observation decentralized supervisor:
first compute a controllable and observable sublanguage
\[K_q := \sup \mathcal {CO}(L_m({\bf SUB}_q));\] and then construct a
{\it partial-observation coodinator}
\begin{align} \label{eq:coor}
{\bf CO}_q = (X_q,\Sigma_q,\eta_q,x_{0,q},X_{m,q})
\end{align}
such that
\begin{align*} 
L_m({\bf SUB}_q) \cap L_m({\bf{CO}}_q) &= K_q\\
L({\bf SUB}_q) \cap L({\bf{CO}}_q) &= \overline{K_q}.
\end{align*}

For the example in Fig.~\ref{fig:modsupsyn}, in ``Step 2",
we decompose the system consisting of four modules into
two subsystems (${\bf SUB}_1$ and ${\bf SUB}_2$), leaving the decentralized supervisor ${\bf SUP}_3$ in between. In case
${\bf SUB}_1$ is blocking (i.e. the two supervisors ${\bf SUP}_1$
and ${\bf SUP}_2$ are conflicting), a partial-observation coordinator ${\bf CO}_1$ is designed to resolve this conflict.

\vspace{1em}
{\it Step 3) Subsystem model abstraction}: After Step~2, the system
consists of a set of nonblocking subsystems. Now we need to verify
the nonconflicting property among these subsystems. For this we use
model abstraction technique with the properties of \emph{natural observer} \cite{FenWon08}
to obtain an abstracted model of each
subsystem,\footnote{The natural observer property of a projection
$P':\Sigma^*\rightarrow \Sigma'^*$
describes that whenever a string $s \in L({\bf G})$ and $P's$
can be extended to $P'L_m({\bf G})$ by an observable string, $s$ can be extended to $L_m({\bf G})$
by the same projection; this property is important for guaranteeing the nonblockingness of the
control design.}  and check the nonconflictingness on the abstractness level,
generally with lower computation complexity. The procedure is as follows:
\begin{enumerate}[(i)]

\item Determine the shared event set, denoted by $\Sigma_{sub}$,
    of these subsystems. Let $P_{sub}:\Sigma^* \rightarrow \Sigma_{sub}^*$ be the corresponding natural projection.


\item For every subsystem check if the corresponding restriction of $P_{sub}$ is an natural observer. If yes,
    let $\Sigma_{sub}' = \Sigma_{sub}$, $P_{sub}'$ be the corresponding natural projection, and goto (iii); otherwise, employ the {\it minimal extension}
    algorithm in \cite{FenWon08} to compute a reasonable extension of $\Sigma_{sub}$ that does define an observer
    for every subsystem. Denote the extended alphabet by $\Sigma_{sub}'$ and
     the corresponding natural projection by $P_{sub}'$.

\item Compute model abstractions for each subsystem with $P_{sub}'$.
\end{enumerate}

Note that there is no particular relationship between $P_{sub}':\Sigma^*\rightarrow \Sigma_{sub}'^*$
and the partial-observation $P$. On the one hand, the projection $P_{sub}'$ guarantees that the control
design at the abstracted level is equivalent to that at the non-abstracted level. On the
other hand, projection $P$ restricts that the control designs at the both levels must
respect to partial-observation. 

This step is illustrated in Fig.~\ref{fig:modsupsyn},
``Step 3", where ${\bf PSUB}_i$ ($i = 1,2$)
with a dashed box denotes the abstraction of subsystem ${\bf SUB}_i$. In addition,
for the intermediate supervisor ${\bf SUP}_3$, we apply the reduction
algorithm \cite{SuWon04} to obtain its (control-equivalent)
reduced model, denoted by ${\bf RSUP}_3$.

\vspace{1em}
{\it Step 4) Abstracted subsystem decomposition and coordination}:
This step is similar to Step 2, but for the abstracted models
instead of modules. We group the abstracted models based on their
interconnection dependencies, and for each group verify the
nonblocking property. If a group turns out to be blocking, we design
a partial-observation coordinator that removes blocking strings.
In Fig.~\ref{fig:modsupsyn}, ``Step 4", we treat the two subsystem abstractions and the intermediate reduced supervisor
as a single group. If this group turns out to be blocking, another coordinator ${\bf CO}_2$ is designed to resolve the conflict.

\vspace{1em}
{\it Step 5) Higher-level abstraction}: Repeat Steps 3 and 4 until
there remains a single group of subsystem abstractions in Step 4.

The heterarchical supervisor/coordinator synthesis terminates at
Step 5; the result is a heterarchical array of partial-observation
decentralized supervisors and coordinators. Specifically, Step 1 gives a set
of partial-observation decentralized supervisors $\{{\bf SUP}_p | p\in \mathcal{P}\}$;
and Step 2 to 5 iteratively generate a set of coordinators, denoted by
$\{{\bf CO}_q | q \in \mathcal{Q}\}$ ($\mathcal{Q}$ an index set).
Similar to \cite{FenWon08}, we prove in Theorem~\ref{thm:equ_large} below that these
partial-observation supervisors/coordinators together achieve globally feasible and
nonblocking (controllable and observable) controlled behavior.

\vspace{1em}
{\it Step 6) Partial-observation localization}: In this last step,
we apply the partial-observation localization algorithm \cite{ZhangCW17} to decompose
each of the obtained decentralized supervisors ${\bf SUP}_p$
($p \in \mathcal{P}$) and coordinators ${\bf CO}_q$ ($q \in \mathcal{Q}$) into
partial-observation local controllers for their corresponding controllable events.
Specifically, for each controllable event
$\alpha \in \Sigma_{c,p}$ ($= \Sigma_c \cap \Sigma_p$), we construct by the partial-observation
localization procedure a partial-observation local controller
${\bf LOC}_{\alpha, p} = (Y_{\alpha,p},\Sigma_{\alpha,p},\eta_{\alpha,p},y_{0,\alpha,p},Y_{m,\alpha,p})$.
By the same procedure, for each ${\bf SUP}_p$, we construct a set of
partial-observation local controllers $\{{\bf LOC}_{\alpha, p}|\alpha \in \Sigma_{c,p}\}$.
Similarly, we localize each ${\bf CO}_q$ to a set of
partial-observation local coordinators $\{{\bf LOC}_{\alpha, q}|\alpha \in \Sigma_{c,q}\}$
where ${\bf LOC}_{\alpha,q} =
(Y_{\alpha,q},\Sigma_{\alpha,q},\eta_{\alpha,q},y_{0,\alpha,q},Y_{m,\alpha,q})$ and $\Sigma_{c,q} = \Sigma_c \cap \Sigma_q$.

\vspace{1em}

We note that the above procedure differs the full-observation one in
\cite{CaiWon10a,CaiWon10b} from: (i) computing {\it partial-observation
decentralized supervisors} and {\it partial-observation coordinators} in Steps 1-5, and (ii)
in Step 6 applying the {\it partial-observation supervisor localization}
developed in Section~III.
By the following Theorem~\ref{thm:equ_large}, the resulting local controllers achieve the
same controlled behavior as the decentralized supervisors and
coordinators did.


\subsection{Main result}


The procedure described above constructs for each controllable event $\alpha$ multiple partial-observation local controllers, because $\alpha$ may belong to different partial-observation decentralized supervisors or coordinators.
In this case, we denote by ${\bf LOC}_\alpha := (X_\alpha, \Sigma_\alpha, \xi_\alpha, x_{0,\alpha}, X_{m,\alpha})$ the synchronous
product of all the local controllers for $\alpha$, i.e.
\begin{align*} 
L({\bf LOC}_\alpha) &= \big(\mathop{||} \limits_{p \in \mathcal{P}} L({\bf LOC}_{\alpha,p})\big) ~||~ \big(\mathop{||} \limits_{q \in \mathcal{Q}} L({\bf LOC}_{\alpha,q})\big)\\
L_m({\bf LOC}_\alpha) &= \big(\mathop{||} \limits_{p \in \mathcal{P}} L_m({\bf LOC}_{\alpha,p})\big) ~||~ \big(\mathop{||} \limits_{q \in \mathcal{Q}} L_m({\bf LOC}_{\alpha,q})\big)
\end{align*}
It can be easily verified that ${\bf LOC}_\alpha$ is also a partial-observation local controller
for $\alpha$, because synchronous product change neither the control authority on $\alpha$
(condition (i)), nor the observation scope $\Sigma_o$ (condition (ii)).

By the same operation (synchronous product) on the partial-observation
local controllers obtained by the localization procedure, we obtain a set
of partial-observation local controllers ${\bf LOC}_\alpha$, one for
each controllable event $\alpha \in \Sigma_c$. We shall verify below that
 these local controllers collectively achieve a safe and nonblocking controlled behavior.

\begin{theorem} \label{thm:equ_large}
The set of partial-observation local controllers $\{{\bf LOC}_\alpha|\alpha \in \Sigma_c\}$
is a solution to the Partial-Observation Supervisor Localization Problem (for large-scale DES), i.e.
\begin{align} %
   &L_m({\bf G})\cap L_m({\bf LOC})\subseteq L_m({\bf G}) \cap P_e^{-1}E \label{eq:sub1:thm}\\
   &L({\bf G})\cap L({\bf LOC})= \overline{L_m({\bf G})\cap L_m({\bf LOC})} \label{eq:sub2:thm}
\end{align}
where $L_m({\bf LOC}) = \mathop \bigcap\limits_{\alpha \in \Sigma_{c}}
   P_\alpha^{-1}L_m({\bf LOC}_{\alpha})$ and $L({\bf LOC}) = \mathop \bigcap\limits_{\alpha \in \Sigma_{c}}
   P_\alpha^{-1}L({\bf LOC}_{\alpha})$
\end{theorem}


This theorem asserts that the local controllers and coordinators
achieve a global nonblocking controlled behavior, that may not be feasibly computable for large-scale
systems in a monolithic way. Instead, by the proposed
heterarchical approach, the partial-observation decentralized supervisors and coordinators are easier
to be obtained, reducing the computational
effort of the localization procedure. This theorem also confirms that
the proposed localization procedure supplies a computable way to
the distributed control problem for large-scale DES under partial
observation; to the best of our knowledge, no result is found in the literature
to deal with this problem.
\vspace{1em}

\noindent {\it Proof of Theorem~\ref{thm:equ_large}}:
The first five steps of the procedure
generate a heterarchical array of partial-observation
decentralized supervisors $\{{\bf SUP}_p | p \in \mathcal{P}\}$ and partial-observation
coordinators $\{{\bf CO}_q | q \in \mathcal{Q}\}$.
We first prove that the collectively controlled behavior
of these decentralized supervisors and coordinators is safe and nonblocking, and then show that the partial-observation
local controllers are control equivalent to the decentralized supervisors
and coordinators.

(i) ({\it safe and nonblocking}) Let $\bf SYS$ represent the collective behavior of these decentralized supervisors
and coordinators, i.e.
\begin{align*}
L_m({\bf SYS}) ~:=~& L_m({\bf G})\cap \big(\bigcap\limits_{p \in \mathcal{P}} P_p^{-1}L_m({\bf SUP}_p)\big) \cap \big( \bigcap \limits_{q \in \mathcal{Q}} P_q^{-1}L_m({\bf CO}_q)\big) \\ 
L({\bf SYS}) ~:=~& L({\bf G})\cap  \big(\bigcap\limits_{p \in \mathcal{P}} P_p^{-1}L({\bf SUP}_p)\big)\cap \big( \bigcap \limits_{q \in \mathcal{Q}} P_q^{-1}L({\bf CO}_q)\big) 
\end{align*}
where $P_p:\Sigma^* \rightarrow \Sigma_p^*$ and $P_q:\Sigma^* \rightarrow \Sigma_q^*$
are the corresponding natural projections.
First, it is easy to verify that $L_m({\bf SYS}) \subseteq L_m({\bf G}) \cap P_e^{-1}E$,
because for each decentralized supervisor, by (\ref{eq:decsup}) $L_m({\bf G}_p) \cap L_m({\bf SUP}_p) = K_p \subseteq  E_p || L_m({\bf G}_p)$
and thus
\begin{align*}
L_m({\bf SYS}) &\subseteq L_m({\bf G}) \cap \big(\bigcap\limits_{p \in \mathcal{P}} P_p^{-1}L_m({\bf SUP}_p)\big)  \\
               &= \bigcap \limits_{p \in \mathcal{P}} P_p^{-1}\big(L_m({\bf G}_p) \cap L_m({\bf SUP}_p)\big) \\
               &\subseteq \bigcap \limits_{p \in \mathcal{P}} P_p^{-1}(E_p || L_m({\bf G}_p)) \\
               & = P_e^{-1}(\mathop{||}\limits_{p \in \mathcal{P}} E_p) \cap L_m({\bf G}) \\
               & = P_e^{-1}E \cap L_m({\bf G})
\end{align*}
Hence, the collective behavior is safe.

Then it follows from \cite[Theorem 4]{FenWon08} that
\[L({\bf SYS}) = \overline{L_m({\bf SYS})}\]
i.e. the collective behavior is nonblocking.

(ii) ({\it control-equivalence})
In Step 6, each decentralized supervisor ${\bf SUP}_p$ ($p \in \mathcal{P}$)
is decomposed into a set of local controllers ${\bf LOC}_{\alpha,p}$, one for each controllable
event $\alpha \in \Sigma_{c,p}$, thus by \cite[Theorem 1]{ZhangCW17},
\begin{align*}
L({\bf G}_p) \cap \big(\mathop{||}\limits_{\alpha \in \Sigma_{c,p}} L({\bf LOC}_{\alpha,p})\big) =& L({\bf G}_p) \cap L({\bf SUP}_p)\\
L_m({\bf G}_p) \cap \big(\mathop{||}\limits_{\alpha \in \Sigma_{c,p}} L_m({\bf LOC}_{\alpha,p})\big) =& L_m({\bf G}_p) \cap L_m({\bf SUP}_p)
\end{align*}

So,
\begin{align*}
L({\bf G}) ~\cap \Big(\bigcap\limits_{\alpha \in \Sigma_{c}}
   P_{\alpha}^{-1} \big(\mathop{||}\limits_{p \in \mathcal{P}}L({\bf LOC}_{\alpha,p})\big)\Big) 
   = &\Big(\mathop \bigcap\limits_{p \in \mathcal{P}}P_p^{-1}L({\bf G}_p)\Big) ~\cap \Big(\bigcap\limits_{p \in \mathcal{P}}
   \big(\mathop{||}\limits_{\alpha \in \Sigma_{c,p}}L({\bf LOC}_{\alpha,p})\big)\Big) \\
   = &\mathop \bigcap\limits_{p \in \mathcal{P}}P_p^{-1}\Big(L({\bf G}_p) ~\cap
   \big(\mathop{||}\limits_{\alpha \in \Sigma_{c,p}}L({\bf LOC}_{\alpha,p})\big)\Big) \\
 = &\mathop \bigcap\limits_{p \in \mathcal{P}}P_p^{-1}\big(L({\bf G}_p) \cap  L({\bf SUP}_p)\big)\\
 = &L({\bf G}) \cap \mathop \bigcap\limits_{p \in \mathcal{P}}P_p^{-1}\big(L({\bf SUP}_p)\big)
\end{align*}
and
\begin{align*}
L_m({\bf G}) ~\cap~ \Big(\bigcap\limits_{\alpha \in \Sigma_{c}}
   P_{\alpha}^{-1} \big(\mathop{||}\limits_{p \in \mathcal{P}}L_m({\bf LOC}_{\alpha,p})\big)\Big) =L_m({\bf G}) \cap \mathop \bigcap\limits_{p \in \mathcal{P}}P_p^{-1}\big(L_m({\bf SUP}_p)\big)
\end{align*}

Similarly, for the coordinators ${\bf CO}_q$ ($q \in \mathcal{Q}$), we
have
\begin{align*}
L({\bf G}) ~\cap~ &\Big(\bigcap\limits_{\alpha \in \Sigma_{c}}
   P_{\alpha}^{-1} \big(\mathop{||}\limits_{q \in \mathcal{Q}}L({\bf LOC}_{\alpha,q})\big)\Big)  =L({\bf G}) \cap \mathop \bigcap\limits_{q \in \mathcal{Q}}P_q^{-1}\big(L({\bf CO}_q)\big) \\
L_m({\bf G}) ~\cap~ &\Big(\bigcap\limits_{\alpha \in \Sigma_{c}}
   P_{\alpha}^{-1} \big(\mathop{||}\limits_{q \in \mathcal{Q}}L_m({\bf LOC}_{\alpha,q})\big)\Big)  =L_m({\bf G}) \cap \mathop \bigcap\limits_{q \in \mathcal{Q}}P_q^{-1}\big(L_m({\bf CO}_q)\big)
\end{align*}

Hence,
\begin{align*}
L({\bf G}) ~\cap ~L({\bf LOC}) 
   = &L({\bf G}) \cap \bigcap\limits_{\alpha \in \Sigma_c}
   P_{\alpha}^{-1}L({\bf LOC}_{\alpha}) \\
   = &\Big[L({\bf G}) ~\cap \Big(\bigcap\limits_{\alpha \in \Sigma_{c}}
   P_{\alpha}^{-1} \big(\mathop{||}\limits_{p \in \mathcal{P}}L({\bf LOC}_{\alpha,p})\big)\Big)\Big]  \cap \Big[L({\bf G}) ~\cap~ \Big(\bigcap\limits_{\alpha \in \Sigma_{c}}
   P_{\alpha}^{-1} \big(\mathop{||}\limits_{q \in \mathcal{Q}}L({\bf LOC}_{\alpha,q})\big)\Big)\Big] \\
   = & L({\bf G})\cap  \big(\bigcap\limits_{p \in \mathcal{P}} P_p^{-1}L({\bf SUP}_p)\big)\cap \big( \bigcap \limits_{q \in \mathcal{Q}} P_q^{-1}L({\bf CO}_q)\big) \\
  =& L({\bf SYS})
\end{align*}
and
\begin{align*}
L_m({\bf G}) ~\cap ~L_m({\bf LOC}) 
   = &L_m({\bf G}) \cap \bigcap\limits_{\alpha \in \Sigma_c}
   P_{\alpha}^{-1}L_m({\bf LOC}_{\alpha}) \\
   = & L_m({\bf G})\cap  \big(\bigcap\limits_{p \in \mathcal{P}} P_p^{-1}L_m({\bf SUP}_p)\big)  \cap \big( \bigcap \limits_{q \in \mathcal{Q}} P_q^{-1}L_m({\bf CO}_q)\big) \\
  =& L_m({\bf SYS})
\end{align*}
which means that the partial-observation local controllers achieve the same controlled
behavior $\bf SYS$ with the decentralized supervisors and coordinators.
By the results in (i), i.e. $\bf SYS$ is safe and nonblocking, the conditions
(\ref{eq:sub1:thm}) and (\ref{eq:sub2:thm}) hold.
\hfill $\square$


\section{Case Study: AGVs} \label{sec:casestudy}

\begin{figure}[!t]
\centering
    \includegraphics[scale = 0.20]{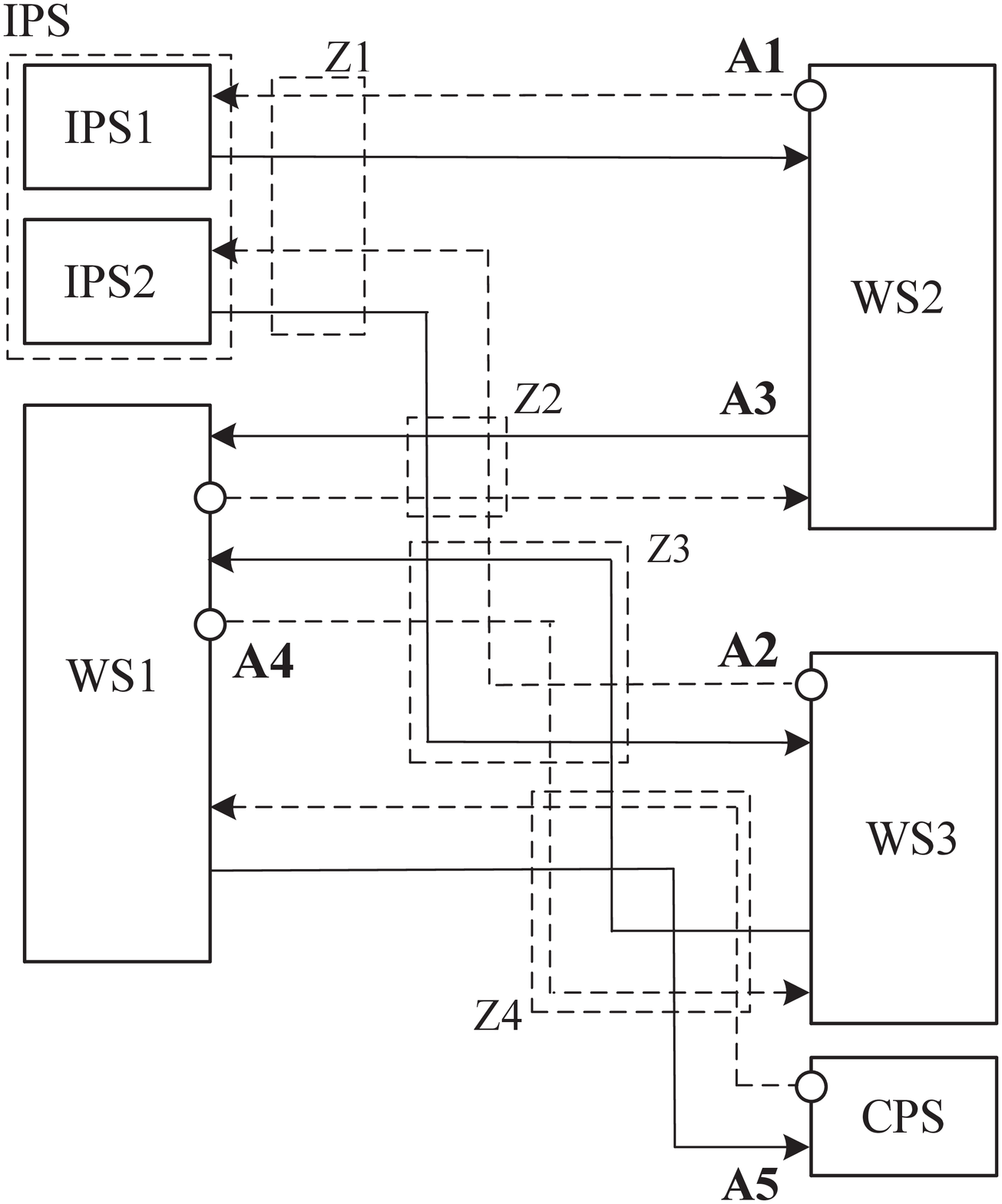}
\caption{AGV system configuration. Rectangular dashed boxes
represent shared zones of the AGV's traveling routes.}
\label{fig:AGVsystem}
\end{figure}
\begin{figure}[!t]
\centering
    \includegraphics[scale = 0.4]{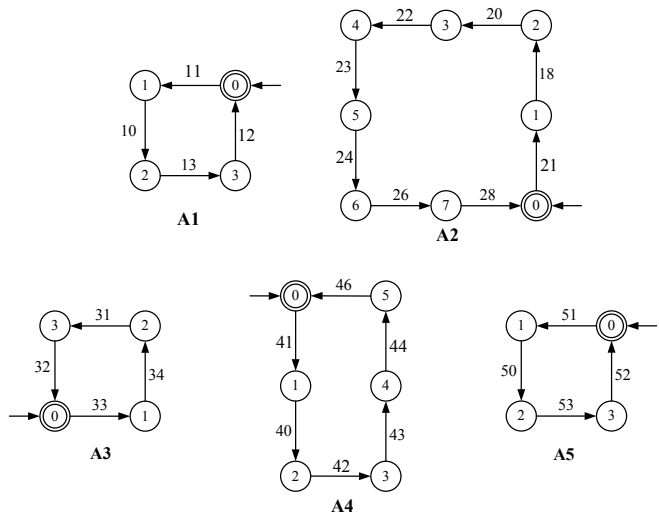}
\caption{AGV: Generators of plant components}
\label{fig:AGVplant}
\end{figure}
\begin{figure}[!t]
\centering
    \includegraphics[scale = 0.4]{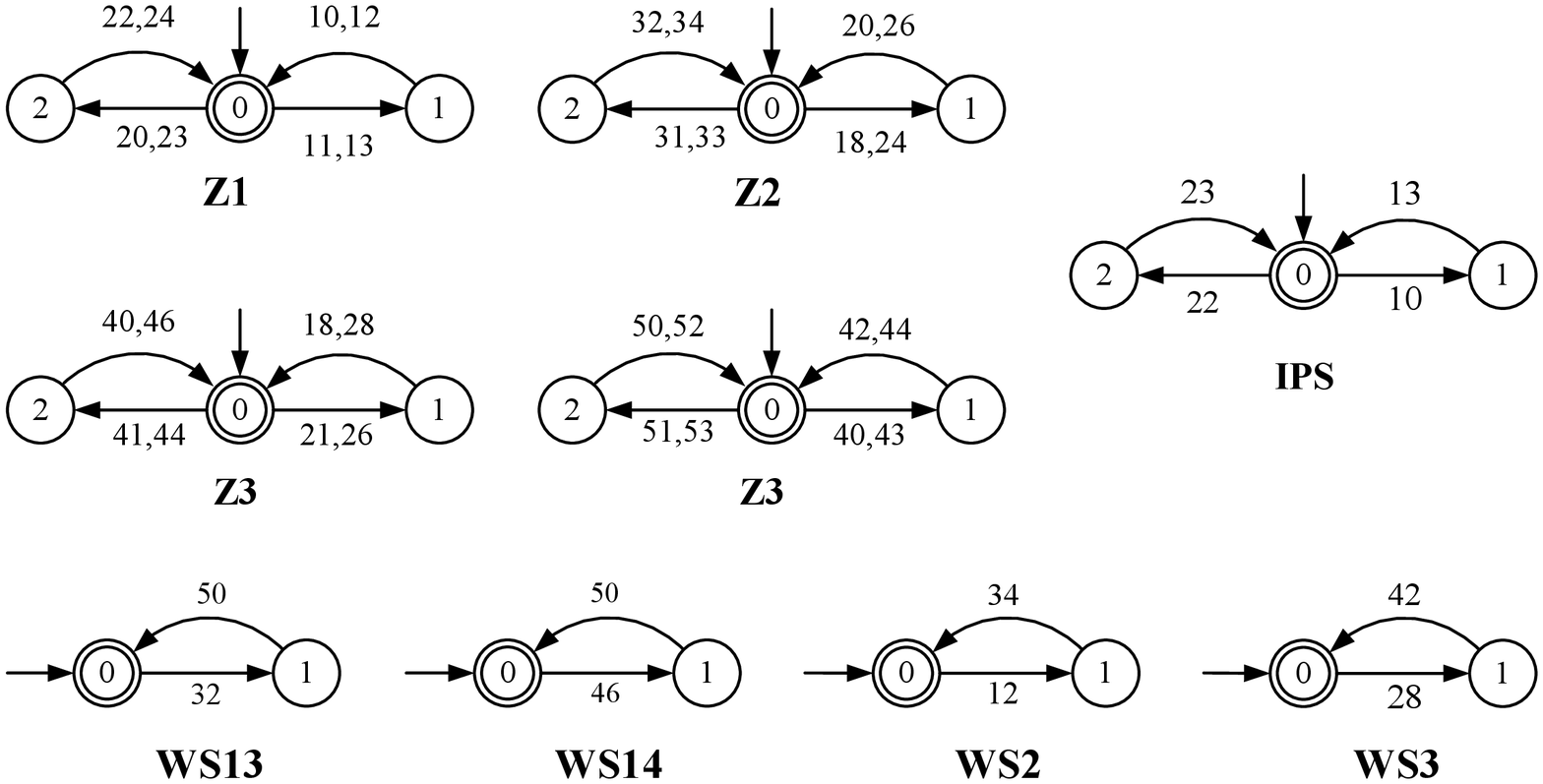}
\caption{AGV: Generators of specifications}
\label{fig:AGVspec}
\end{figure}

In this section we apply the proposed heterarchical localization procedure
to study the distributed control of AGV serving a manufacturing
workcell under partial observation. As displayed in
Fig.~\ref{fig:AGVsystem}, the plant consists of five independent AGV
\[{\bf A1}, {\bf A2}, {\bf A3}, {\bf A4}, {\bf A5}\]
and there are nine imposed control specifications
\begin{align*}
{\bf Z1}, {\bf Z2}, {\bf Z3}, {\bf Z3},{\bf WS13}, {\bf WS14S}, {\bf WS2}, {\bf WS3}, {\bf IPS}
\end{align*}
which require no collision of AGV in the shared zones and no
overflow or underflow of buffers in the workstations. The generator
models of the plant components and the specification are displayed in
Figs.~\ref{fig:AGVplant} and \ref{fig:AGVspec} respectively; the
detailed system description and the interpretation of the events are
referred to \cite[Section 4.7]{Wonham16a}.

\begin{table}
\footnotesize
\caption{Physical interpretation of unobservable events} \label{tab:unobservEvent}
\begin{center}
\scalebox{0.8}{
\begin{tabular}{|c||c|}
\hline
Event & Physical interpretation \\
\hline
$13$ & $\bf A1$ re-enters Zone 1   \\
\hline
$23$ & $\bf A2$ re-enters Zone 1   \\
\hline
$31$ & $\bf A3$ re-enters Zone 2   \\
\hline
$42$ & $\bf A4$ exists Zone 4 and loads from WS3   \\
\hline
$53$ & $\bf A5$ re-enters Zone 4   \\
\hline
\end{tabular}
}
\end{center}
\end{table}

Consider partial observation and let the unobservable event set be
$\Sigma_{uo} = \{13, 23, 31, 42, 53\}$; thus each AGV has an unobservable
event and the corresponding physical interpretation is listed in Table~\ref{tab:unobservEvent}.
Our control objective is to design for each AGV a set of local strategies
subject to partial observation such that
the overall system behavior satisfies the imposed specifications and is nonblocking.

\begin{figure}[!t]
\centering
    \includegraphics[scale = 0.38]{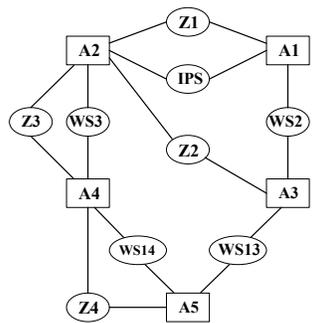}
\caption{Event-coupling relations}
\label{fig:event_couple1}
\end{figure}

\vspace{1em}
{\it Step 1) Partial-observation decentralized supervisor synthesis}: For each specification
displayed in Fig.~\ref{fig:AGVspec}, we group its event-coupled AGV as the decentralized plant (see
Fig.~\ref{fig:event_couple1}), and synthesize as in (\ref{eq:decsup}) a partial-observation
decentralized supervisor. The state sizes of these decentralized supervisors are displayed
in Table~\ref{tab:sup&coor}, in which the supervisors
are named correspondingly to the specifications, e.g. $\bf Z1SUP$
is the decentralized supervisor corresponding to the specification $\bf Z1$.

\begin{table}
\footnotesize
\caption{State sizes of partial-observation decentralized supervisors} \label{tab:sup&coor}
\begin{center}
\scalebox{0.8}{
\begin{tabular}{|c|c||c|c|}
\hline
Supervisor & State size & Supervisor & State size\\
\hline
$\bf Z1SUP$ & 13 & $\bf Z2SUP$ & 11  \\
\hline
$\bf Z3SUP$ & 26 & $\bf Z4SUP$ & 9\\
\hline
$\bf WS13SUP$ & 15 & $\bf WS14SUP$ & 19\\
\hline
$\bf WS2SUP$ & 15 & $\bf WS3SUP$ & 26\\
\hline
$\bf IPSSUP$ & 13 &  &\\
\hline
\end{tabular}
}
\end{center}
\end{table}

\begin{figure}[!t]
\centering
    \includegraphics[scale = 0.14]{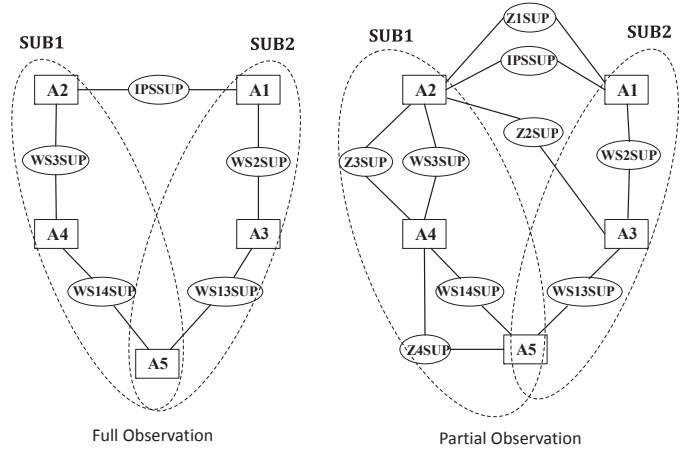}
\caption{Subsystem decomposition}
\label{fig:InterConnect}
\end{figure}

\vspace{1em}
{\it Step 2) Subsystem decomposition and coordination:} We have nine
decentralized supervisors, and thus nine modules (consisting of a decentralized
supervisor with associated AGV components). Under full observation, the
decentralized supervisors for the four zones ($\bf Z1SUP$, ..., $\bf Z4SUP$)
are {\it harmless} to the overall nonblocking property \cite[Proposition 5]{FenWon06},
and thus can be safely removed from the interconnection structure;
then the interconnection structure of these modules are simplified by applying
{\it control-flow net} \cite{FenWon08}. Under partial observation,
however, the four decentralized supervisors are not harmless to the overall
nonblocking property (also by \cite[Proposition 5]{FenWon06}, the necessary conditions
are not satisifed due to partial observation) and thus cannot be removed. As displayed in Fig.~\ref{fig:InterConnect},
we decompose the overall system into two subsystems:
\begin{align*}
{\bf SUB1} := &{\bf A2} || {\bf A4} || {\bf A5} ||  {\bf WS3SUP} || {\bf WS14SUP} || {\bf Z3SUP} || {\bf Z4SUP}\\
{\bf SUB2} := &{\bf A1} || {\bf A3} || {\bf A5} ||  {\bf WS2SUP} || {\bf WS13SUP}
\end{align*}
Between the two subsystems are decentralized supervisors $\bf Z1SUP$, $\bf Z2SUP$, and $\bf IPSSUP$.
It is verified that $\bf SUB2$ is nonblocking, but $\bf SUB1$ is blocking. Hence we design a
coordinator $\bf CO1$ (as in (\ref{eq:coor})) which makes $\bf SUB1$ nonblocking.
This coordinator $\bf CO1$ has 36 states, and we refer to this nonblocking subsystem $\bf NSUB1$.

\begin{figure}[!t]
\centering
    \includegraphics[scale = 0.18]{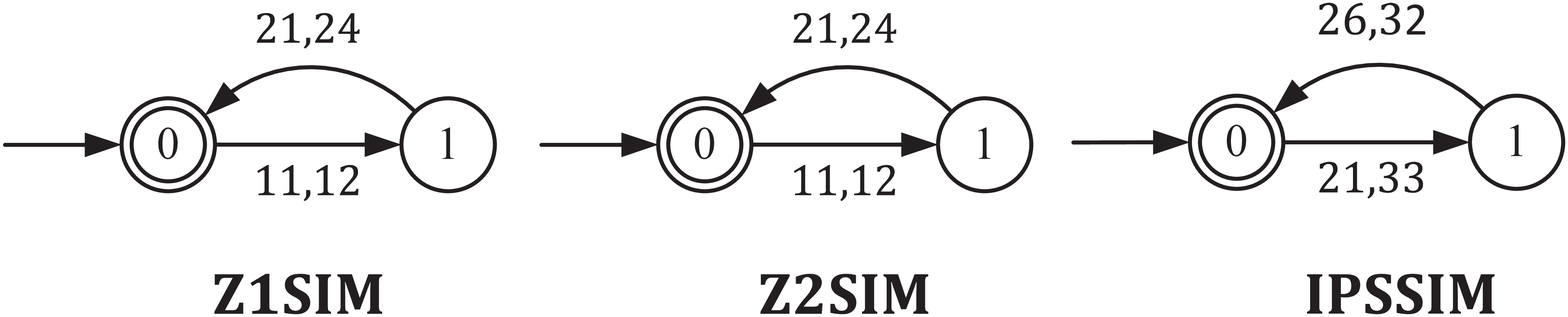}
\caption{Reduced generator models of decentralized supervisors ${\bf Z1SUP}$, ${\bf Z2SUP}$ and ${\bf IPSSUP}$}
\label{fig:simsup}
\end{figure}

\vspace{1em}
{\it Step 3) Subsystem model abstraction}: Now we need to verify the
nonconflicting property among the nonblocking subsystems ${\bf NSUB1}$, ${\bf SUB2}$ and the decentralized
supervisors ${\bf IPSSUP}, {\bf Z1SUP}$ and ${\bf Z2SUP}$.
First, we determine their shared event set, denoted by $\Sigma_{sub}$.
Subsystems ${\bf NSUB1}$ and ${\bf SUB2}$ share all events in $\bf A5$:
50, 51, 52 and 53. For ${\bf IPSSUP}, {\bf Z1SUP}$ and ${\bf Z2SUP}$, we use
their reduced generator models ${\bf IPSSIM}$, ${\bf Z1SIM}$ and $\bf Z2SIM$
by supervisor reduction \cite{SuWon04}, as displayed in Fig.~\ref{fig:simsup}.  By inspection, ${\bf IPSSUP}$ and ${\bf Z1SIM}$
share events 21 and 24 with ${\bf NSUB1}$, and events 11 with ${\bf SUB2}$;
${\bf Z2SUP}$ shares events 24 and 26 with ${\bf NSUB1}$, and events 32, 33 with ${\bf SUB2}$.
Thus \[\Sigma_{sub} = \{11,12,21,24,26,32,33,50,51,52,53\}.\]
It is then verified that $P_{sub}:\Sigma^* \rightarrow \Sigma_{sub}^*$
satisfies the natural observer property \cite{FenWon08}. With $P_{sub}$, therefore,
we obtain the subsystem model abstractions, denoted by ${\bf QC\_NSUB1} = P_{sub}({\bf NSUB1})$ and
${\bf QC\_SUB2} = P_{sub}({\bf SUB2})$, with state sizes listed in Table~\ref{tab:subsys}.
\begin{table}
\footnotesize
\caption{State sizes of model abstractions} \label{tab:subsys}
\begin{center}
\scalebox{0.8}{
\begin{tabular}{|c||l|l|}
\hline
 & $\bf NSUB1$ ~~ ${\bf QC\_NSUB1}$ & $\bf SUB2$ ~~ $\bf QC\_SUB2$\\
\hline
State size & ~~~50 ~~~~~~~~~~~~~ 19 & ~~~574 ~~~~~~~~~ 56  \\
\hline
\end{tabular}
}
\end{center}
\end{table}

\vspace{1em}
{\it Step 4) Abstracted subsystem decomposition and coordination}: We treat ${\bf QC\_NSUB1}$,
${\bf QC\_SUB2}$, $\bf IPSSIM$, $\bf Z1SIM$ and $\bf Z2SIM$ as a single group, and check the nonblocking property.
This group turns out to be blocking, and a coordinator $\bf CO2$ is then designed (as in (\ref{eq:coor})) 
to make the group nonblocking. This coordinator $\bf CO2$ has 123 states.

%
%
%

\vspace{1em}
{\it Step 5) Higher-level abstraction}: The modular supervisory control
design terminates with the previous Step 4.

%
%
We have obtained a heterarchical array of nine partial-observation decentralized supervisors and
two coordinators. These supervisors and coordinators together achieve a globally feasible
and nonblocking controlled behavior.

\begin{figure}[!t]
\centering
    \includegraphics[scale = 0.16]{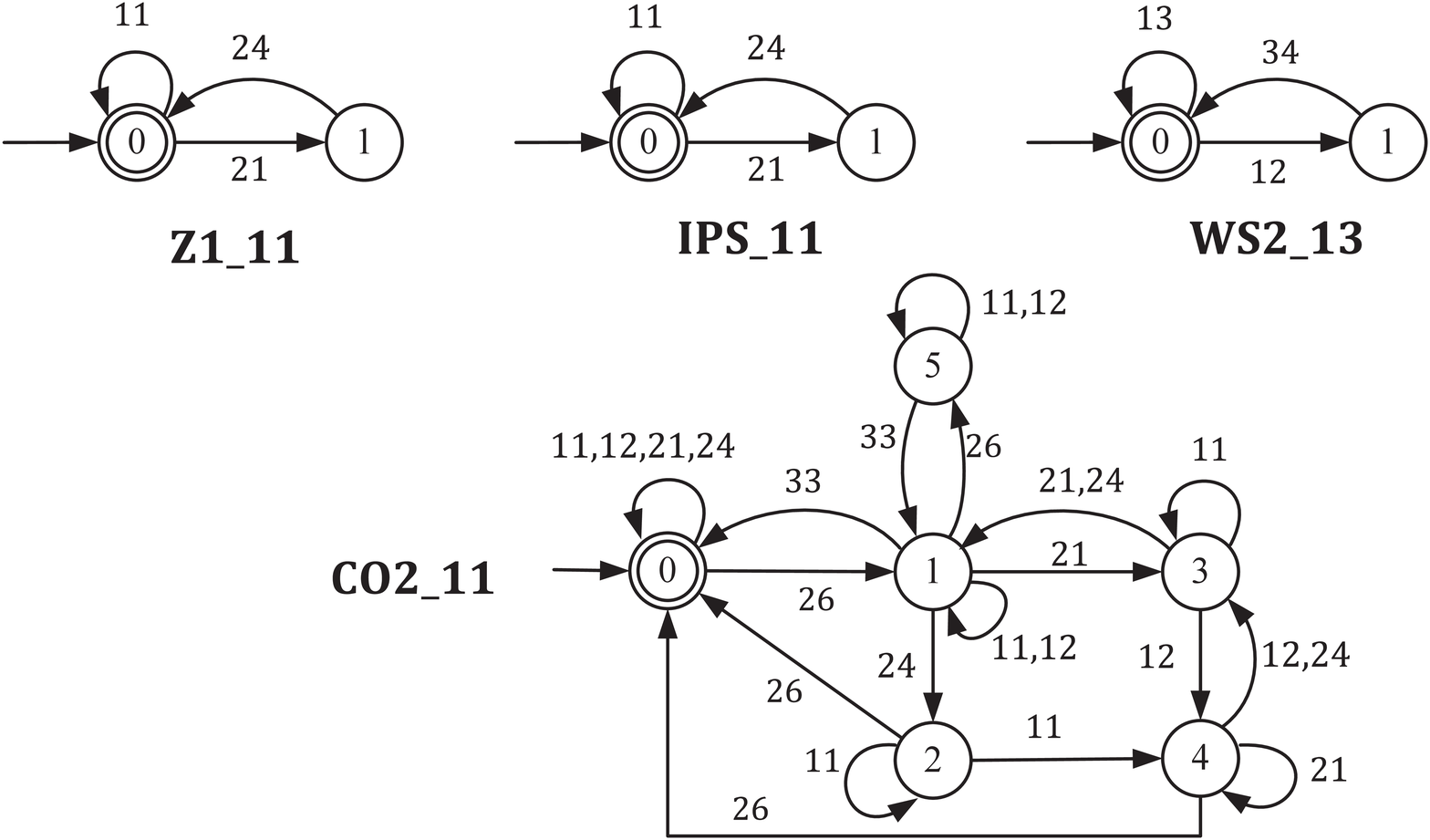}
\caption{Partial-observation local controllers and coordinators for $\bf A1$ with controllable events 11 and 13 (the local controllers
are named in the format of `specification\_event')}
\label{fig:LOC_A1}
\end{figure}

\begin{figure}[!t]
\centering
    \includegraphics[scale = 0.4]{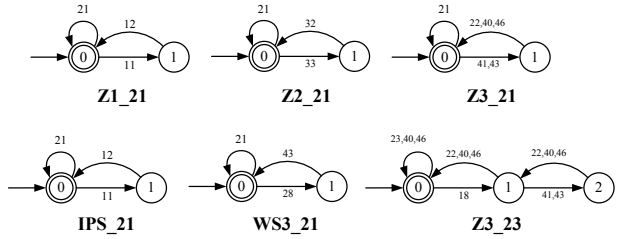}
\caption{Partial-observation local controllers for $\bf A2$ with controllable events 21 and 23}
\label{fig:LOC_A2}
\end{figure}

\begin{figure}[!t]
\centering
    \includegraphics[scale = 0.16]{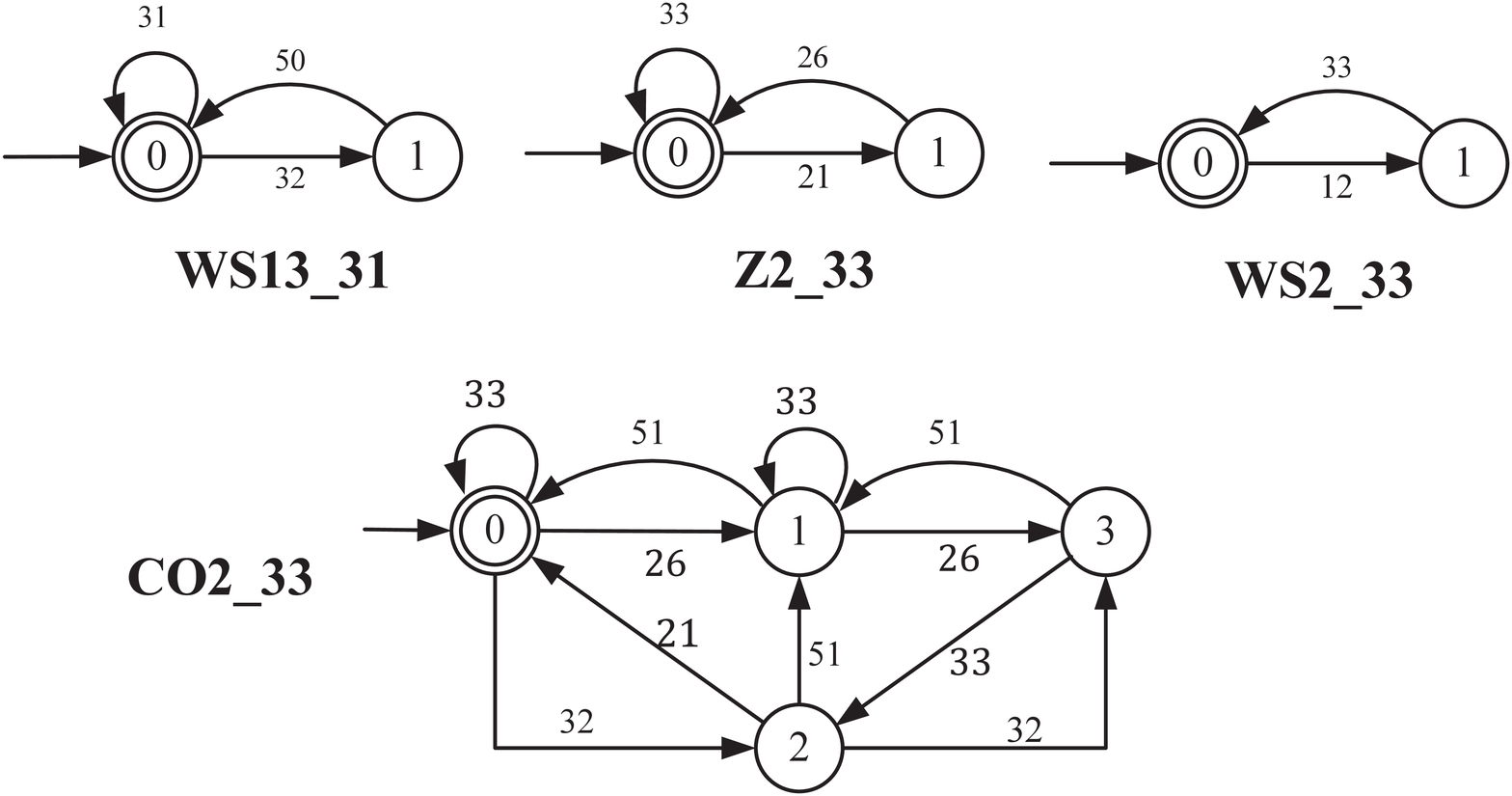}
\caption{Partial-observation local controllers and coordinators for $\bf A3$ with controllable events 31 and 33}
\label{fig:LOC_A3}
\end{figure}

\begin{figure}[!t]
\centering
    \includegraphics[scale = 0.14]{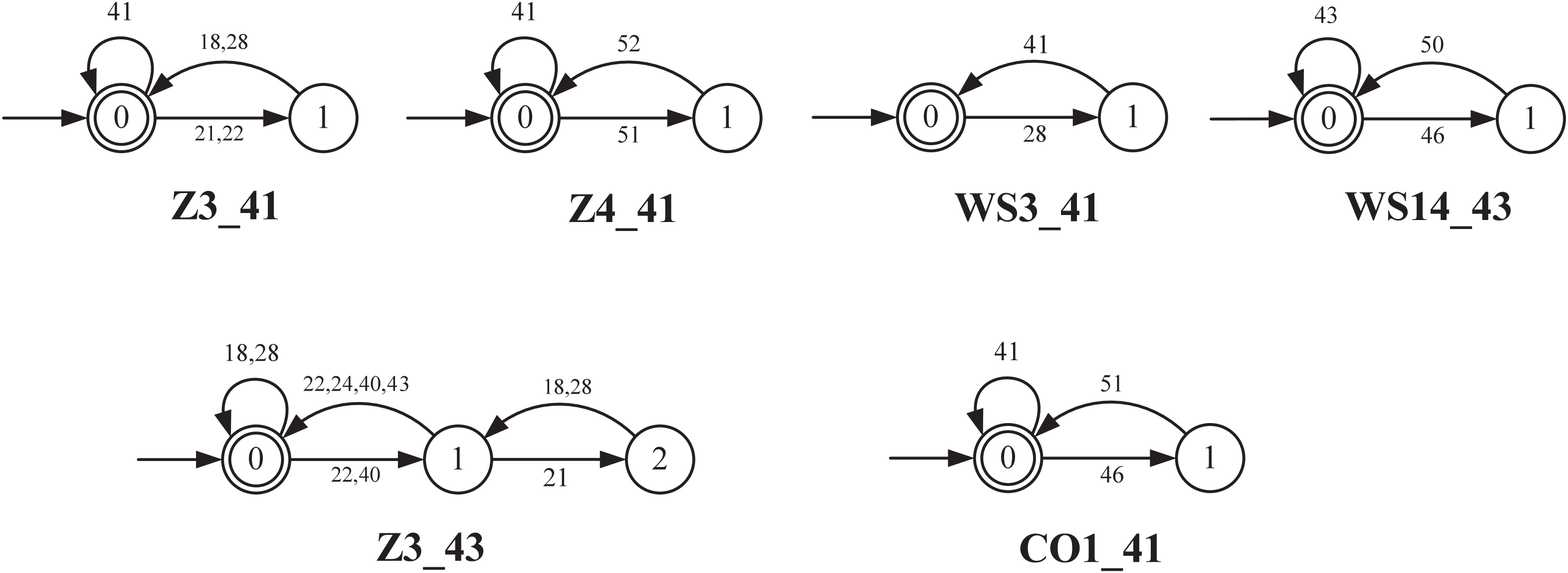}
\caption{Partial-observation local controllers and coordinators for $\bf A4$ with controllable events 41 and 43}
\label{fig:LOC_A4}
\end{figure}

\begin{figure}[!t]
\centering
    \includegraphics[scale = 0.4]{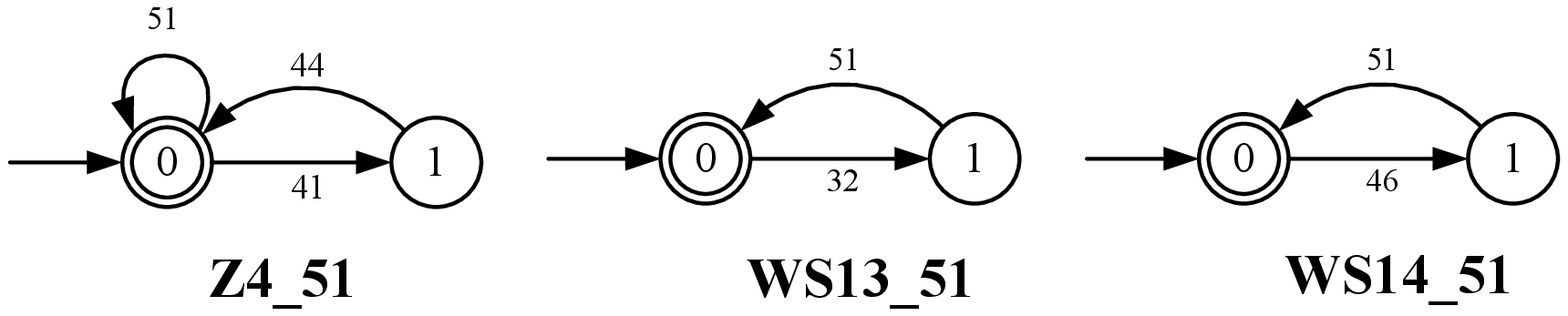}
\caption{Partial-observation local controllers for $\bf A5$ with controllable events 51 and 53 (event 53 is not disabled
and thus there is no corresponding local controller)}
\label{fig:LOC_A5}
\end{figure}

\vspace{1em}
{\it Step 6) Partial-observation localization}: We finally apply the partial-observation supervisor
localization procedure \cite{ZhangCW17} to decompose the obtained decentralized
supervisors/coordinators into local controllers/coordinators under partial observation.
The generator models of
the local controllers/coordinators are displayed in Fig.~\ref{fig:LOC_A1}-\ref{fig:LOC_A5};
they are grouped with respect to the individual AGV and their state sizes
are listed in Table~\ref{tab:loc}. By inspecting the transition structures
of the local controllers/coordinators, only observable events lead to states changes.


\begin{table*}
\footnotesize
\caption{State sizes of partial-observation local controllers/coordinators}
\label{tab:loc}
\begin{center}
\scalebox{0.72}{
\begin{tabular}{|c||c|c|c|c|c|}
\hline  & Local controller/coordinator of & Local controller of & Local
controller/coordinator of & Local controller/coordinator of & Local controller of
\\
Supervisor/coordinator & $\bf A1 (state~ size)$ & $\bf A2 (state~
size)$ & $\bf A3 (state~ size)$
& $\bf A4 (state ~size)$& $\bf A5 (state ~size)$\\
\hline
$\bf Z1SUP$ & $\bf Z1\_11 (2)$ & $\bf Z1\_21 (2)$&&&\\
\hline
$\bf Z2SUP$ & &$\bf Z2\_21 (2)$&$\bf Z2\_33 (2)$&&\\
\hline
$\bf Z3SUP$ &&$\bf Z3\_21 (2)$,$\bf Z3\_23 (3)$&&$\bf Z3\_41 (2)$,$\bf Z3\_43 (3)$&\\
\hline
$\bf Z4SUP$ & &&&$\bf Z4\_41 (2)$&$\bf Z4\_51 (2)$\\
\hline
$\bf WS13SUP$ & && $\bf WS13\_31 (2)$&&$\bf WS13\_51 (2)$\\
\hline
$\bf WS14SUP$ & &&&$\bf WS14\_43 (2)$&$\bf WS14\_51 (2)$\\
\hline
$\bf WS2SUP$ & $\bf WS2\_13 (2)$&&$\bf WS2\_33 (2)$&&\\
\hline
$\bf WS3SUP$ & &$\bf WS3\_21 (2)$&&$\bf WS3\_41 (2)$&\\
\hline
$\bf IPSSUP$ & $\bf IPS\_11 (2)$&$\bf IPS\_21 (2)$&&&\\
\hline
$\bf CO1$ & &&&$\bf CO1\_41 (2)$&\\
\hline
$\bf CO2$ & $\bf CO2\_11 (6)$&&$\bf CO2\_33 (4)$&&\\
\hline
\end{tabular}
}
\end{center}
\end{table*}

Partial observation affects the control logics of the controllers/coordinators
and thus affects the controlled system behavior. For illustration, consider the
following case: assuming that event sequence 11.10.13.12.21.18.20.22 has
occurred, namely $\bf A1$ has loaded a type 1 part to workstation
$\bf WS2$, and $\bf A2$ has moved to input station $\bf IPS2$. Now,
$\bf A2$ may load a type 2 part from $\bf IPS2$ (namely, event 23 may
occur). Since event 24 ($\bf A2$ exits Zone 1 and re-enter Zone 2) is
uncontrollable, to prevent the specification on Zone 2 ($\bf Z2$) not
being violated, AGV $\bf A3$ cannot enter Zone 2 if 23 has occurred,
i.e. event 33 must be disabled. However, event 33 is eligible to occur
if event 23 has occurred. So, under the full observation condition
(event 23 is observable) event 33 would occur safely if event 23 has
not occurred. However the fact is that event 23 is unobservable;
so due to (relative) observability, 33 must also be disabled even if
23 has not occurred, namely the controllers will not know whether or
not event 23 has occurred, so it will disabled event 33 in both cases,
to prevent the possible illegal behavior. This control strategy coincides
with local controller $\bf Z2\_33$: event 33 must be disabled if event
21 has occurred, and will not be re-enabled until event 26 has occurred
($\bf A2$ exits Zone 2 and re-enter Zone 3).

\begin{figure}[!t]
\centering
    \includegraphics[scale = 0.13]{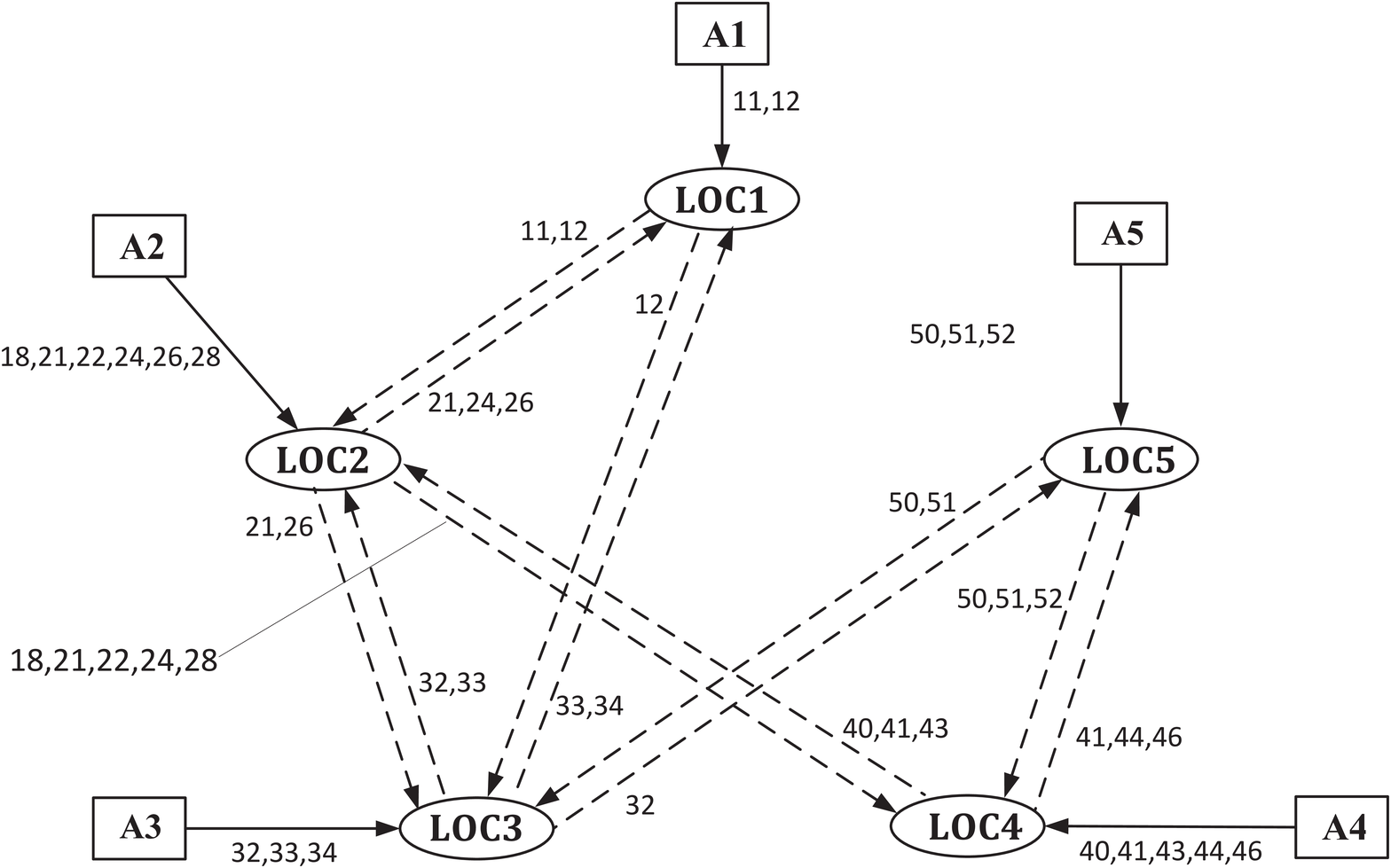}
\caption{AGV: communication diagram of local controllers/coordinators. For $i = 1,...,5$,
${\bf LOCi}$ represents the local controllers/coordinators corresponding to ${\bf Ai}$. } \label{fig:AGV_com}
\end{figure}

Finally, the heterarchical supervisor localization has effectively
generated a set of partial-observation local controllers/coordinators with small
state sizes (between 2 and 6 states). Grouping these local
controllers/coordinators for the relevant AGV, we obtain a distributed control
architecture for the system where each AGV is controlled by its own
controllers while observing certain observable events of other AGV;
according to the transition diagrams of the local controllers/coordinators, we obtain a
communication diagram, as displayed in Fig.~\ref{fig:AGV_com}, which shows the events to
be observed (denoted by solid lines) or communicated (denoted
by dashed lines) to local controllers/coordinators.

\section{Conclusions} \label{sec:concl}
We have developed a systematic top-down approach to
solve the distributed control of large-scale multi-agent DES under partial
observation. This approach first employs relative observability and
an efficient heterarchical synthesis procedure to
compute a heterarchical array of partial-observation
decentralized supervisors and partial-observation coordinators, and then
decomposes the decentralized supervisor/coordinators
into a set of partial-observation local controllers whose state
changes are caused only by observable events. Moreover,
we have proved that these local controllers collectively
achieve a globally nonblocking behavior.
An AGV example has been presented for illustration.

\small
\bibliographystyle{IEEEtran}
\bibliography{SCDES_Ref}

\end{document}